\documentclass[twocolumn,aps,prl]{revtex4-1}
\usepackage{graphicx}
\usepackage{dcolumn}
\usepackage{bm}

\begin{document}
\newcommand*{\bi}{\bibitem}
\newcommand*{\ea}{\textit{et al.}}
\newcommand*{\eg}{\textit{e.g.}}
\newcommand*{\zpc}[3]{Z.~Phys.~C \textbf{#1}, #2 (#3)}
\newcommand*{\plb}[3]{Phys.~Lett.~B \textbf{#1}, #2 (#3)}
\newcommand*{\phrd}[3]{Phys.~Rev.~D~\textbf{#1}, #2 (#3)}
\newcommand*{\phrl}[3]{Phys.~Rev.~Lett.~\textbf{#1}, #2 (#3)}
\newcommand*{\pr}[3]{Phys.~Rev.~\textbf{#1}, #2 (#3)}      
\newcommand*{\npbps}[3]{Nucl.~Phys.~B (Proc. Suppl.) \textbf{#1}, #2 (#3)}  
\newcommand*{\ibid}[3]{\textit{ ibid.} \textbf{#1}, #2 (#3)}
\newcommand*{\epjc}[3]{Eur. Phys. J. C \textbf{#1}, #2 (#3)}
\newcommand*{\ra}{\rightarrow}
\newcommand*{\tauch}{\tau^-\ra\,\nu_\tau\,\pi^-\pi^+\pi^-}
\newcommand*{\taunn}{\tau^-\ra\,\nu_\tau\,\pi^-\pi^0\pi^0}
\newcommand*{\tripi}{\pi^-\pi^+\pi^-}
\newcommand*{\pitwo}{\pi^-\pi^0\pi^0}
\newcommand*{\amu}{{\mathbf A}^\mu}
\newcommand*{\anu}{{\mathbf A}^\nu}
\newcommand*{\amunu}{{\mathbf A}^{\mu\nu}}
\newcommand*{\vmu}{{\mathbf V}_\mu}
\newcommand*{\vmunu}{{\mathbf V}_{\mu\nu}}
\newcommand*{\fai}{\mathbf P}
\newcommand*{\rf}[1]{(\ref{#1})}
\newcommand*{\lag}{{\mathcal L}}
\newcommand*{\sth}{\sin\theta}
\newcommand*{\be}{\begin{equation}}
\newcommand*{\ee}{\end{equation}}
\newcommand*{\arp}{{a_1\rho\pi}}
\newcommand*{\bea}{\begin{eqnarray}}
\newcommand*{\eea}{\end{eqnarray}}
\newcommand*{\nl}{\nonumber \\}
\newcommand*{\mas}{m_{a_1}^2}
\newcommand*{\aone}{a_1(1260)}
\newcommand*{\apr}{a_1^\prime}
\newcommand*{\aonep}{a_1(1640)}
\newcommand*{\aonepp}{a_1(1420)}

\title{Resonance $\mathbf{a_1(1420)}$ and the Three-Pion Decays of the Tauon}

\author{Peter Lichard}
\affiliation{
Institute of Physics, Silesian University in Opava, 746 01 Opava, 
Czech Republic\\
and\\
Institute of Experimental and Applied Physics, Czech
Technical University in Prague, 128 00 Prague, Czech Republic
}
\begin{abstract}
The role of the $\aonepp$ resonance in the three-pion 
decays of the $\tau$ lepton is investigated using a phenomenological
model. For all data before 2008, roughly equal fit quality  is
achieved when the basic $\aone$ resonance is supplemented with either $\aonep$
or $\aonepp$. However,  two recent and more precise data sets require 
resonances with masses that are not very far from that of $\aonepp$.
This suggests that the axial-vector resonance that accompanies 
the $\aone$ in the three-pion decays of the tauon is $\aonepp$, not 
$\aonep$, as believed up to now. More data are needed to demonstrate this 
definitely.
\end{abstract}
\maketitle
Recently, I have compared our model \cite{vojlic} of the three-pion decays 
of the $\tau$
lepton to nineteen experimental data sets found in literature from 1986 to 
2013. A detailed account will be presented elsewhere \cite{lichardprep}. 

In this Letter I report on one interesting byproduct
of my study that is related to the $\aonepp$ resonance, which was discovered 
in 2013 by the COMPASS Collaboration at CERN \cite{compassprl}. They
performed a careful partial wave analysis of the $\pi^-\pi^-\pi^+$ system
produced in diffractive dissociation of 190~GeV$/c$ pions off a stationary
hydrogen target and then used the resonance-model fit to the spin density
matrix to find the resonance parameters.

\begin{figure}[h]
\includegraphics[width=0.485\textwidth,height=0.21\textwidth]{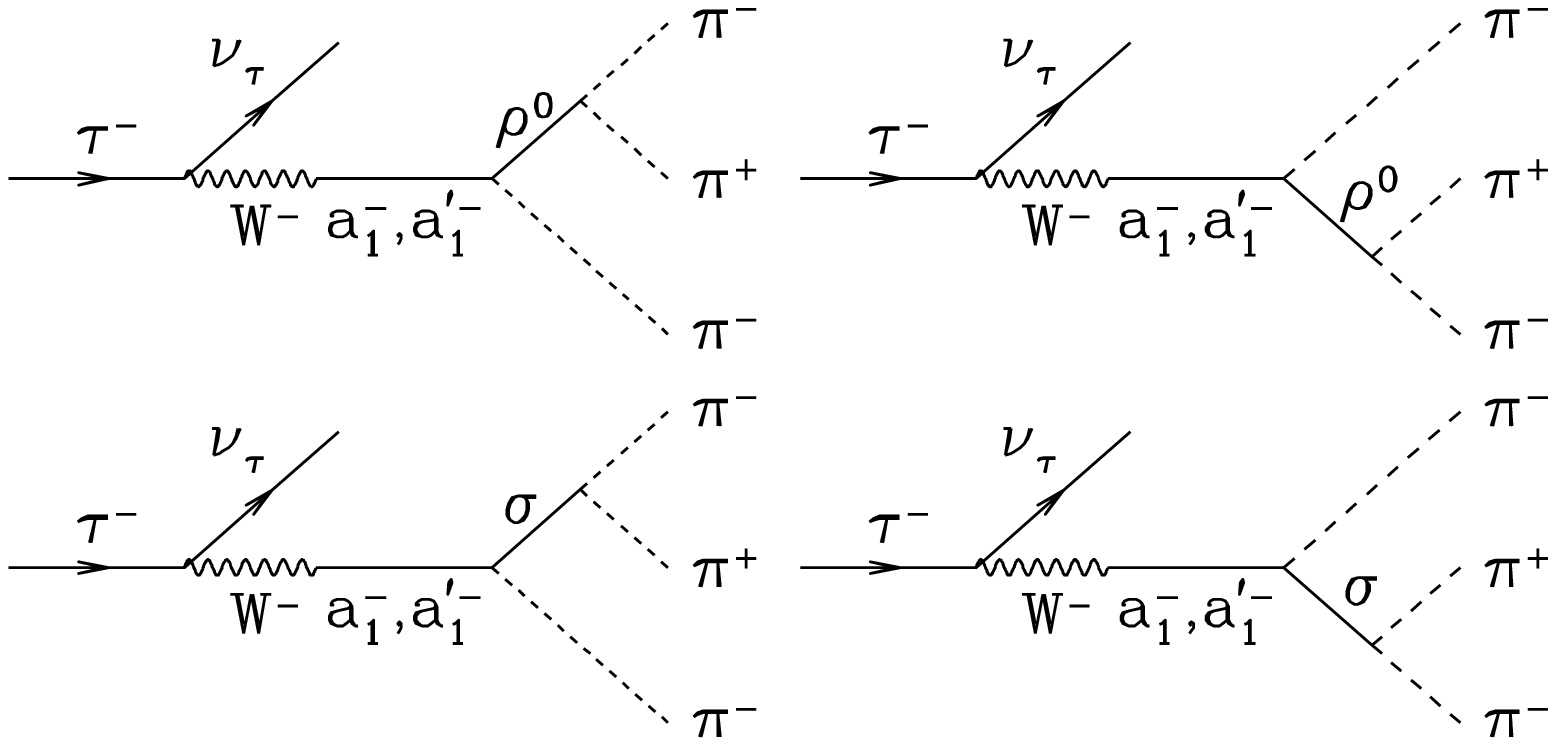}
\caption{\label{fig:fig1}Feynman diagrams of the $\tauch$ decay.}
\end{figure}

\begin{figure}[h]
\includegraphics[width=0.485\textwidth,height=0.21\textwidth]{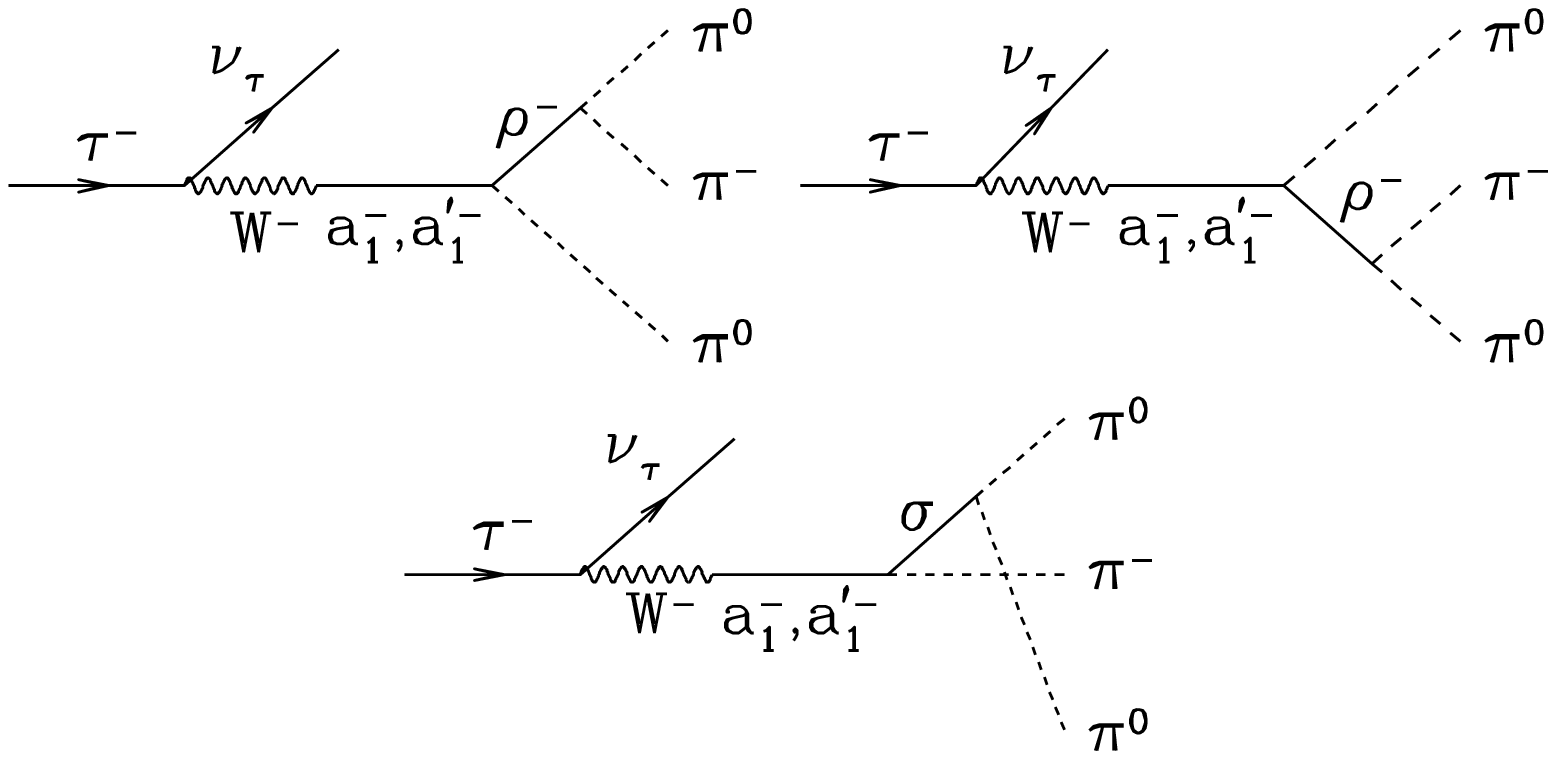}
\caption{\label{fig:fig2}Feynman diagrams of the $\taunn$ decay.}
\end{figure}

I first recapitulate the main features of our model \cite{vojlic}. Its physical 
content is depicted by the meson dominance \cite{vavd} diagrams in Figs. 
\ref{fig:fig1} and \ref{fig:fig2}. 

The interaction among the $\aone$ (denoted simply as $a_1$ in what
follows), $\rho$ and $\pi$ fields is described by a two-component Lagrangian 
\be
\label{lagarp}
\lag_\arp=\frac{g_{\arp}}{\sqrt{2}}
\left(\lag_1\cos\theta+\lag_2\sin\theta\right),
\ee
where
$\lag_1 = \amu\cdot\left(\vmunu\times\partial^\nu{\fai}\right)$,
$\lag_2 = \vmunu\cdot\left(\partial^\mu\anu\times{\fai}\right)$,
and $\vmunu = \partial_\mu{\mathbf V}_\nu-\partial_\nu\vmu$.
The isovectors $\amu$, $\vmu$, and $\fai$ denote the operators of 
the $a_1$, $\rho$ and $\pi$ fields, respectively. Although one can find
several flaws in this Lagrangian from the QCD point of view, its soundness
for low-energy phenomenology is supported by the fact that the same
Lagrangian provided a good description of the four-pion production in the
$e^+e^-$ annihilation \cite{licjur}, even for the same value of the mixing
parameter $\sth$, as shown in \cite{vojlic}. 

The $a_1$ propagator 
\be
\label{a1prop}
-i G^{\mu\nu}_{a_1}(p) = \frac{-g^{\mu\nu}+p^\mu p^\nu/\mas}
{s-M^2_{a_1}(s)+im_{a_1}\Gamma_{a_1}(s)}
\ee
features running mass $M_{a_1}(s)$ given by a once-subtracted dispersion 
relation with the energy-dependent total width $\Gamma_{a_1}(s)$ as input. 
The following contributions to $\Gamma_{a_1}(s)$ are considered:
$a_1 \ra \rho+\pi\ra 3\pi$, $a_1 \ra \bar{K^*}K, K^*\bar{K} 
\ra K\bar{K}\pi,$ and $a_1 \ra \sigma+\pi\ra 3\pi$ and the following 
conditions are satisfied: 
$M^2_{a_1}(\mas)=\mas$,
$\Gamma_{a_1}(\mas)=\Gamma_{a_1}$.

When an additional axial-vector resonance $\apr$ is included, the same 
Lagrangian \rf{lagarp} is used  with the same mixing parameter $\sth$.
Instead of \rf{a1prop}, a simpler $\apr$ propagator is chosen, with running 
mass replaced by the nominal $\apr$ mass and with the energy dependent total 
width given by 
$
\Gamma_{\apr}(s)=k\Gamma_{a_1}(s),
$
with constant $k$ guaranteeing that the condition $\Gamma_{\apr}(m^2_{\apr})=
\Gamma_{\apr}$ is fulfilled.

The propagator of the $\rho(770)$ resonance is taken in a variable-width,
running-mass form \cite{runningrho}.

The interaction Lagrangian among the $a_1$, $\pi$, and $\sigma$ 
fields is written in the form
\be
\lag_{a_1\sigma\pi}=g_1\left(\amu\cdot\partial_\mu{\fai}\right)S+g_2
\left(\amu\cdot\fai\right)\partial_\mu S,
\ee
where $S$ is the operator of the $\sigma$ field. For the $\sigma$ propagator
we use a simple form with the fixed mass and energy dependent width.

All strong interaction vertexes are modified by the form factor 
$F(q)=\exp\{-q^2/(12\beta^2)\}$,
where $q$ is the three-momentum magnitude of a daughter meson in the
rest frame of the parent meson. It is taken from the chromoelectric 
flux-tube breaking model of Kokoski and Isgur \cite{kokoski1987},
together with their value $\beta=0.4$~GeV/$c$.

The model without $\apr$ contains six adjustable parameters: (i) a 
multiplicative constant ensuring the correct normalization to the data, 
which absorbs the product of all coupling constants from the $a_1\ra\rho$ 
branch, (ii) $a_1$ mass, (iii) $a_1$ width, (iv) $\arp$ Lagrangian 
mixing parameter $\sth$, (v-vi) two different products of the coupling 
constants from the $a_1\ra\sigma$ branch scaled by the product of coupling 
constants from the $a_1\ra\rho$ branch. 

If an $\apr$ is included, a complex constant enters, which multiplies 
the $\apr$ propagator before adding it to the $a_1$ propagator. 
The number of adjustable parameters is thus eight. If the mass and width 
of $\apr$ are allowed to vary, then the number of adjustable parameters 
rises to ten. Finally, if a combined 
set of $N$ individual data  sets is fitted, then the number of adjustable 
parameters increases by $N\!-\!1$ because every data set requires its own 
normalization constant.

When we were completing our model \cite{vojlic}, the existing data on the
three-pion decays of the tauon were not precise enough to consider the 
$\aonep$ mass and width as free parameters. Therefore, we used their values 
from the 2008 Review of Particle Physics \cite{pdg2008}, which are still
used today. As I will show, the situation has changed with the advent of 
two more precise sets of data. The first, which appeared in the PhD Thesis 
of I. M. Nugent \cite{nugent09}, published in 2009, may be regarded as a very 
preliminary version of the BaBar data \cite{note1}. The other set contains 
very precise data from the Belle Collaboration at the KEKB collider, published 
in 2010 \cite{belle10}. 

An important indicator of the soundness of our model is that the parameters
required for a good fit to various data sets are very similar. It is
therefore possible to obtain a satisfactory  fit for several data sets 
combined together, as shown in \cite{vojlic}. 

The original model \cite{vojlic} has been slightly modified: (i) the bin width
in the modified ALEPH data \cite{aleph13} varies from point to point
in the high mass part of the spectrum, which 
required a corresponding change in the software; (ii) following the advice of 
Dr. David Bugg \cite{bugg}, I introduced an option to mimic the Adler zero 
\cite{adler} in the $\sigma\ra\pi+\pi$ amplitude; (iii) in the original 
model \cite{vojlic}, one of the two $a_1\sigma\pi$ coupling constants was 
fixed by requiring the vanishing derivative of the $a_1$ running mass at the 
resonance point. I have released this connection and introduced another 
free parameter (already included in the list above).

After I learned of the discovery of the $\aonepp$ resonance by the COMPASS 
Collaboration \cite{compassprl}, I made all the calculations with 
both options, namely with $\apr\equiv\aonep$ and $\apr\equiv\aonepp$. 
The results presented here were evaluated with the $\aonepp$ mass of 1411.8 MeV  
and width of 158 MeV in accordance with a recent COMPASS submission to arXiv 
\cite{krinner}. 

Concerning the influence of the presence or choice of a particular 
$\apr$ on the agreement of our model with data, the experimental data can 
be divided into three categories.

In the first category, there are four data sets \cite{note2}
(CELLO 90c \cite{cello90}, CELLO 90m \cite{cello90}, 
Nugent 09c \cite{nugent09}, Belle 10c \cite{belle10}) which our model is not 
able to describe (confidence level, C.L., was less than 10\%) whether the 
$\apr$ is included or not. 

Then, there is a group of eight data sets (Argus 93c \cite{argus93}, 
OPAL 95c \cite{opal95}, OPAL 97c \cite{opal97}, ALEPH 98m \cite{aleph98}, 
OPAL 99c \cite{opal99}, OPAL 99m \cite{opal99}, ALEPH 05m \cite{aleph05}, 
ALEPH 13m \cite{aleph13}), five of them with C.L. of 100\%, with which our 
model agrees, even if the $\apr$ is not considered.

The last category includes seven experiments (Mark II 86c \cite{markii86}, 
MAC 87c \cite{mac87}, MAC 87m \cite{mac87},
ALEPH 98c \cite{aleph98}, CLEO 00m \cite{cleo00}, ALEPH 05c \cite{aleph05}, 
ALEPH 13c \cite{aleph13}) in which the inclusion of the $\apr$ in the model 
improves its agreement with data (for three of them, the C.L. then reached 
more than 98\%). The quality of the fit depends only marginally on the $\apr$ 
species.

The examples of data sets from each category are shown in Table
\ref{tab:comparison}. The results of five selected individual data sets are
accompanied by the result of a simultaneous fit to six data sets 
(ARGUS 93c, OPAL 99c, OPAL 99m, CLEO 00m, ALEPH 05c, ALEPH 05m), which is 
denoted as Set A. The first two lines (Set A and ALEPH 13m)
show examples of a good fit achieved without the help of $\aonep$ or
$\aonepp$. If either of them is included, the $\chi^2$ decreases, but there
is no influence on the already perfect confidence level. 
\begin{table*}[h]
\caption{\label{tab:comparison}%
Confidence level (C.L.) of the fits to various data assuming various 
axial-vector recurrences $\apr$. 
Usual $\chi^2$ and the number of degrees of freedom (NDF) are also shown.}
\begin{ruledtabular}
\begin{tabular}{lcdrrcdrrcdrr}
&~&\multicolumn{3}{c}{No $\apr$}&~~~~~~~&\multicolumn{3}{c}{$\apr\equiv\aonep$}
&~~~~~~~&\multicolumn{3}{c}{$\apr\equiv\aonepp$}\\
Data &
&\mbox{$\chi^2$}& \text{NDF} &\text{C.L.}(\%)&
&\mbox{$\chi^2$}& \text{NDF} &\text{C.L.}(\%)&
&\mbox{$\chi^2$}& \text{NDF} &\text{C.L.}(\%)\\
\colrule
Set A   && 368.0 & 462 & 100.0 && 267.1 & 460 & 100.0 && 300.6 & 460 & 100.0\\
ALEPH 13m && 31.2 & 68 & 100.0 && 30.3 & 66 & 100.0 && 30.5 & 66 & 100.0 \\
ALEPH 05c && 102.9 & 110 & 67.0&&  23.7 & 108 & 100.0 && 45.2 & 108 & 100.0\\
ALEPH 13c && 103.6 & 67& 0.3 && 73.6 & 65 & 21.7 && 74.3 & 65 & 20.1 \\
Nugent 09c && 552.8 & 59 & 0.0 && 194.1 & 57 & 0.0 && 78.8 & 57 & 2.9 \\
Belle 10c && 1607.1 & 126 & 0.0 && 845.4 & 126& 0.0 && 572.5& 126 & 0.0
\end{tabular}
\end{ruledtabular}
\end{table*}

Sets ALEPH 05c and ALEPH 13c are examples of data that are fitted 
better when any $\apr$ is included. For ALEPH 13c, the confidence level 
rose from 0.3 to 21.7 per cent [20.1 per cent] if $\aonep$ [$\aonepp$] is
chosen. For ALEPH 05c, confidence level even reaches 100 per cent for either
choice.The last two rows show that our model is unable to fit Nugent 09c 
and Belle 10c data even if $\apr$ is considered.

\begin{figure}
\includegraphics[width=0.485\textwidth]{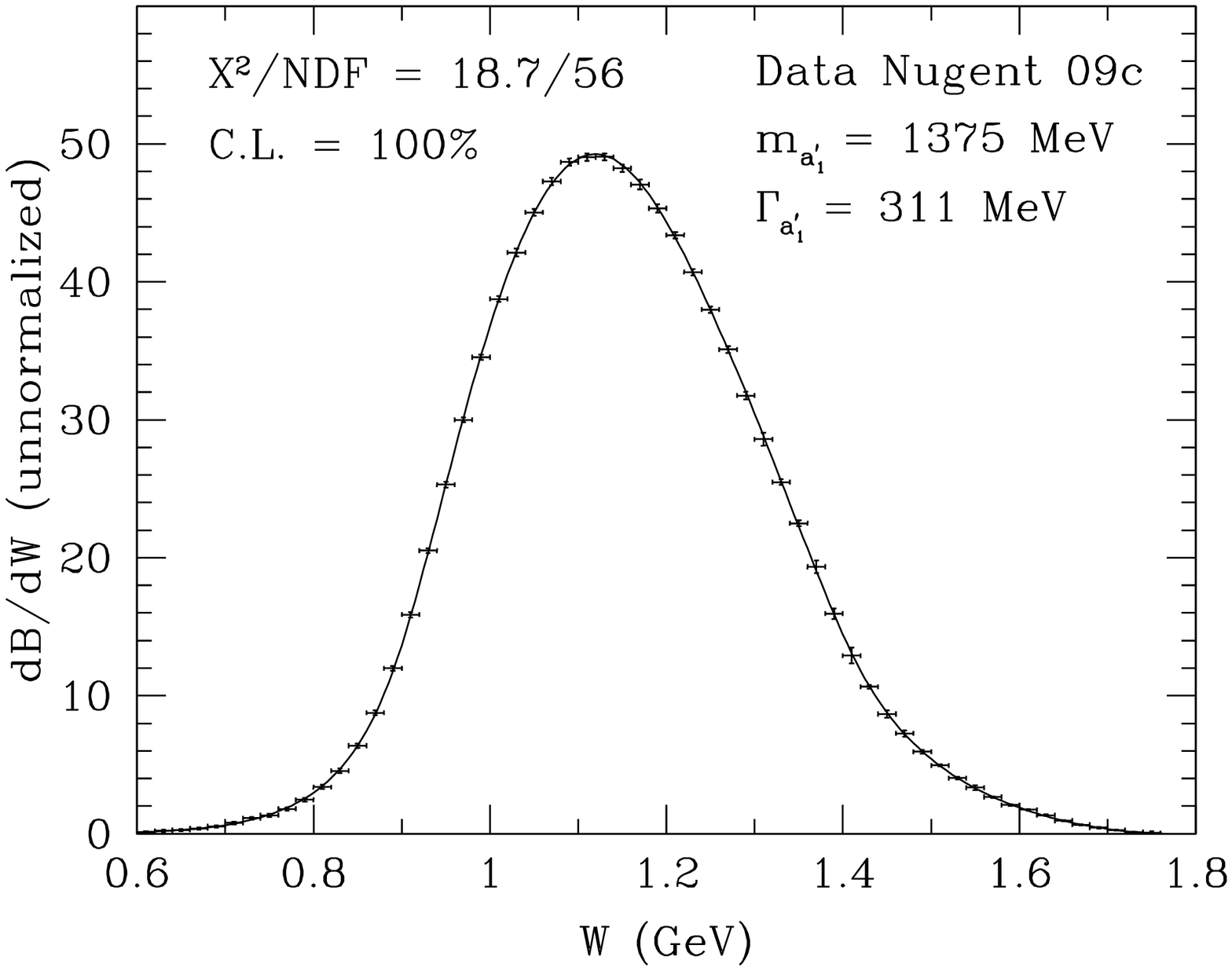}
\caption{\label{fig:nugent}Differential branching fraction in the invariant
mass of the $\tripi$ system. Comparison of the model with the Nugent 09c data
\cite{nugent09}.}
\end{figure}

In any case, it is encouraging that with the Nugent 09c data, replacing 
the $\aonep$ with $\aonepp$ leads to a dramatic drop in the $\chi^2$ 
and some increase in the confidence level, from $\chi^2/$NDF= 194.1/57 
(C.L.=0\%) to $\chi^2/$NDF=78.8/57 (C.L.=2.9\%). This has prompted me
to allow both the $a_1^\prime$ mass and width vary. As a result, 
the confidence level climbs to 100\%, based on $\chi^2/$NDF= 18.7/55. 
Corresponding values of the $a_1^\prime$ mass and width are shown in 
Table \ref{tab:search}, together with their Minuit \cite{minuit} errors. 
A comparison of the experimental and calculated three-pion mass spectra 
appears in Fig. \ref{fig:nugent}.

\begin{figure}
\includegraphics[width=0.485\textwidth]{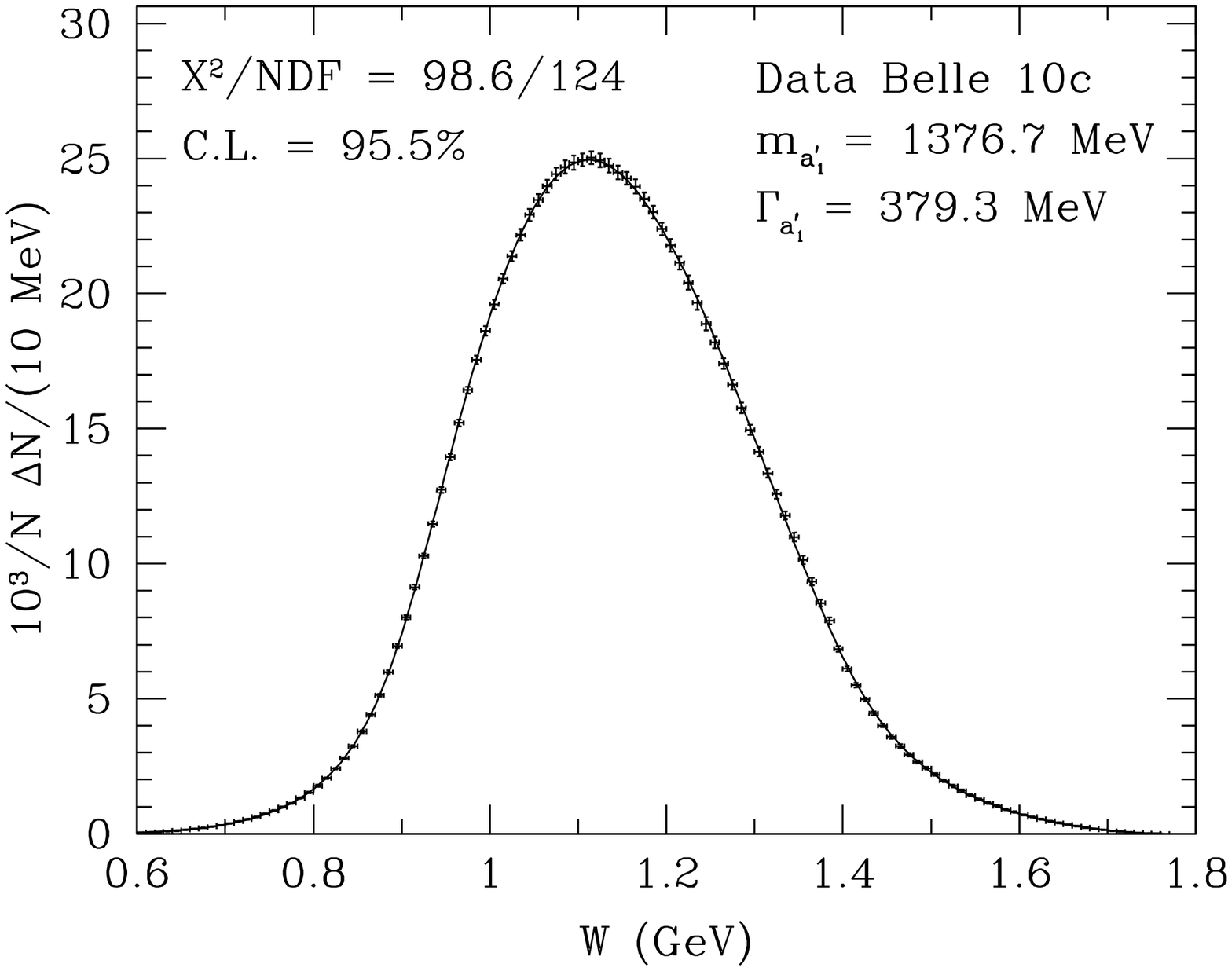}
\caption{\label{fig:belle}Three-pion mass spectrum measured by the Belle
Collaboration \cite{belle10} fitted by our model.}
\end{figure}

Repeating the procedure with the Belle 10c data, I have again achieved a very
good fit (C.L.=95.5\%) for the parameters displayed in Table \ref{tab:search}.
The $\apr$ mass is almost identical to that of the Nugent 09c data; though the
width is higher. The model curve goes perfectly through the experimental 
points, which have very small errors. See Fig. \ref{fig:belle}.

It may seem disturbing that the resonances that have been found
in the Nugent 09c and Belle 10c data by minimizing $\chi^2$ are not
visible in Figs. \ref{fig:nugent} and \ref{fig:belle} 
as bumps or shoulders. 
The reason is that the interference between the $\aone$ and $\apr$ is
destructive, as discussed in \cite{vojlic}. See Figs. 5 and 6 there. 

\begin{table}[h]
\caption{\label{tab:search}%
Searching for the $\apr$ parameters. Only errors coming from Minuit
\cite{minuit} are shown, no attempt to estimate the systematic errors
of the model has been made. Also the COMPASS \cite{compassprl} results 
are shown for comparison.}
\begin{ruledtabular}
\begin{tabular}{lccccrc}
Data&$m$ (MeV)& $\Gamma$ (MeV) &\mbox{$\chi^2$}&\text{NDF}&\text{C.L.}(\%)\\
\colrule
Nugent 09c&$1375\pm11$ & $311\pm8$ & 18.7 & 56 & 100.0 \\
Belle 10c & $1376.7\pm1.2$ & $379.3\pm6.3$ & 98.6 & 124 & 95.5\\
\colrule
COMPASS & $1414^{+15}_{-13}$ & 
$153^{+~8}_{-23}$ & & &
\end{tabular}
\end{ruledtabular}
\end{table}

\begin{table}[h]
\caption{\label{tab:aleph13c}%
Fit to the ALEPH 13c \cite{aleph13} data assuming the $\apr$ parameters
found from the Nugent 09c and Belle 10c data.}
\begin{ruledtabular}
\begin{tabular}{lccc}
 &\mbox{$\chi^2$}&\text{NDF}&\text{C.L.}(\%)\\
\colrule
$\apr$ from Nugent 09c& 65.1 & 65 & 47.3 \\
$\apr$ from Belle 10c & 65.9 & 65 & 44.6
\end{tabular}
\end{ruledtabular}
\end{table}

The situation with the ALEPH 13c data is special. An attempt to obtain the 
optimal $\apr$ parameters by minimizing $\chi^2$ was unsuccessful. The 
minimization procedure strayed into an unphysical region. But when 
the $\apr$ parameters extracted from the Nugent 09c or Belle 10c data
shown in Table \ref{tab:search} were used, the resulting confidence level 
became much better than that with the $\aonep$ or $\aonepp$. Compare Table
\ref{tab:aleph13c} with ALEPH 13c row in Table \ref{tab:comparison}. 

It is tempting to assume that the resonance extracted from the Nugent 09c and 
Belle 10c data, which is also preferred by the ALEPH 13c data, is identical to
the resonance $\aonepp$ observed by the COMPASS Collaboration. It would be
surprising to have two different resonances so close in mass.

However, when we consult Table \ref{tab:search} and compare the parameters of 
the $\apr$ resonance from the study of the three-pion decays of the
tauon with those found by the COMPASS Collaboration, we find a serious 
discrepancy in the decay widths. The $\aonepp$ decay width is about half
of those from the Nugent 09c and Belle 10c data. A possible
reason for this discrepancy may be a different parametrization of the
resonance width. The COMPASS Collaboration used the width modified by the 
Blatt--Weisskopf centrifugal-barrier factors in their relativistic
Breit--Wigner formulas \cite{haas}. On the other hand, in our model
the $\apr$ width is obtained by rescaling the width of $\aone$. 

The COMPASS Collaboration has shown that the $\aonepp$ decays into pion 
and $f_0(980)$. The main drawback of our model is that we consider the 
$\sigma\equiv f_0(500)$ instead.  However, at this stage, it does not make
much sense to begin a labor intensive modification of our model by allowing 
different scalar resonances for $\aone$ and $\aonepp$. Our model in its 
present form shows perfect agreement with both the Nugent 09c and Belle 10c 
data. Clearly, more data are needed that would falsify our model and force 
it to discriminate between the $\sigma$ and $f_0(980)$.

To conclude, the phenomenological analysis of the Nugent \cite{nugent09} and
Belle \cite{belle10} data offers a serious hint that the excited axial-vector
resonance  which plays a role in the three-pion decays of the tauon is the
$\aonepp$, not $\aonep$.

\begin{acknowledgments}
Earlier collaboration with Dr. Martin Voj\'{\i}k is gratefully acknowledged.
Martin independently coded the model \cite{vojlic} in \textsc{C++} programming
language, and by reaching the same results as from my Fortran code, provided 
an important test of the software. 
Thanks also to  Drs. Mikihiko Nakao (for Belle spokespersons) and MyeongJae Lee 
for promptly providing the Belle data, to Dr. David Bugg for 
valuable correspondence, and to Dr. Josef Jur\'{a}\v{n} for discussions.
This work was partly supported by the Czech MSMT projects Inter-Excellence No. 
LTT17018 and INGO II No. LG15052.
\end{acknowledgments}

\end{document}